\documentclass[epj,draft]{svjour}

\begin{document}
\title{
Thermodynamic properties of the Dicke model in the strong-coupling
regime}
\author{Giuseppe Liberti\mail{liberti@fis.unical.it} and Rosa Letizia
Zaffino}
\institute{
Dipartimento di Fisica, Universit\`a della Calabria\\
INFN - Gruppo collegato di Cosenza, 87036 Rende (CS) Italy}
\date{\today}
\bigskip
\abstract{We discuss the problem of a N two-level systems
interacting with a single radiation mode in the strong-coupling
regime. The thermodynamic properties of Dicke model are analyzed
developing a perturbative expansion of the partition function in
the high-temperature limit and we use this method to investigate
the connections between the Dicke and the collective
one-dimensional Ising model.} \PACS{ {42.50.Fx}{Cooperative
phenomena in quantum optical systems }\and {05.70.Jk}{Critical
point phenomena} \and {73.43.Nq}{Quantum phase transition}}
\authorrunning{G. Liberti and R. L. Zaffino}
\titlerunning{Thermodynamic properties of the Dicke model in the strong-coupling
regime} \maketitle

\section{\label{intro}Introduction}
The thermodynamic properties of the Dicke model \cite{dicke} for N
two-level system interacting with a single mode of the radiation
field have been studied extensively by different authors and with
different methods
\cite{hepp,WH,vertogen,Duncan,gilmore,Yamanoi,Orszag,widom}.
Interest in this model has been renewed by a number of theoretical
works, where it is discussed in connection with quantum chaos
\cite{Emary0,Emary1} and entanglement \cite{Hou,lambert,reslen}
and for various physical systems as photonic band gap materials
\cite{quang,lee} and Josephson junctions \cite{frasca}.

The Dicke model exhibits a second-order phase transition and,
after the first derivation due to Hepp and Lieb \cite{hepp}, a
simple computational method was provided by Wang and Hioe
\cite{WH} based on the use of Glauber's coherent states
\cite{glauber} for the radiation field and on the assumption that,
in the thermodynamic limit ($N\rightarrow \infty$, $V\rightarrow
\infty$ but $\rho=N/V$ finite), the field operators can be treated
as $c$-number functions. The Wang and Hioe method has been
recently applied by Lee and Johnson \cite{lee,lee2} to derive the
thermodynamic properties for an extended version of the Dicke
Hamiltonian incorporating spin-spin and spin-boson interactions.
Perturbative methods have recently been proposed to perform
thermodynamical calculations without to have recourse to the Wang
and Hioe computational method. In Ref. \cite{liberti} it has been
developed a perturbative expansion of partition function and a
simple analytic solution is found for high coupling constant. The
existence of this soluble model allows to show that the
interaction of independent atomic spins with a resonant photon
mode can be interpreted as an effective spin-spin interaction of
long range nature. The similarity between the Dicke Model and the
collective XY-model has been found in Ref. \cite{reslen}, using
another perturbative approach in order to derive a
temperature-dependent effective atomic hamiltonian. The prediction
of a phase transition has been criticized as obtained for
nonrealistic Hamiltonian, the simple Dicke Hamiltonian in which
both the counter rotating terms and the diamagnetic term are
truncated, that violated gauge invariance. As a matter of fact,
the critical properties do not changes qualitatively when the
counter rotating terms are taken into account \cite{Duncan} but,
when the diamagnetic term is included, the Super-radiant Phase
Transition (SPT) is forbidden to occur \cite{birula}. The same
conclusion has been obtained in Ref. \cite{liberti} for a model
Hamiltonian incorporating the effects of the diamagnetic term into
a new frequency of the photon mode \cite{crisp}.

In this paper we present a formal analysis which generalizes the
results of Ref. \cite{liberti} beyond the resonant condition and
gives an alternative derivation of some already known and recent
results on the subject. In the next section we analyze the
procedure that permit to develop a perturbative expansion of the
partition function as generally found in literature. The Dicke
Model is introduced in Sec. \ref{modela} and a simple expression
for the partition function is analytically derived in Sec.
\ref{partition} and \ref{field} for high value of the coupling
constant. This general result, as shown in App. \ref{A}, is
independent of the choice for the basis for expressing the state
of radiation field. Some special cases of interest are explored
and the salient features of the thermodynamic phase transition are
discussed in Sec. \ref{phase} and in Sec. \ref{modelb}, where we
also discuss the analogies between the Dicke model and the
collective one-dimensional Ising model. In Sec. \ref{conclu} we
draw our final conclusions.
\section{\label{model}Perturbation expansion of partition function}
The starting point for deriving the thermodynamic properties of a
quantum model defined by an Hamiltonian $H$ is the partition
function:
\begin{equation}
    Z(N,T)={\mathrm{Tr}}\left\{e^{-\beta
    H}\right\}
    \label{zeta0}
    \end{equation}
It is generally difficult to diagonalize $e^{-\beta H}$ and a
variety of systematic approximations has been proposed
\cite{approx1,approx2,approx3}. A convenient procedure is to
introduce a separation
\begin{equation}
    H=H_{0}+H_{I}
    \label{hsplit}
    \end{equation}
such that the exponential operator of (\ref{zeta0}) can be
disentangled into a product of an infinite series of exponential
operators as
\begin{equation}
    e^{-\beta (H_{0}+H_{I})}=e^{-\beta H_{0}}e^{-\beta
    H_{I}}\prod_{i=2}^\infty e^{(-\beta)^{i} C_{i}}
    \label{Zasse}
    \end{equation}
where $C_{i}$ is a homogeneous polynomial of degree $n$ in $H_{0}$
and $H_{I}$. All the $C_{i}$${}^{'}$s contain the commutator
$[H_{I},H_{0}]$ and using the method given by Wilcox \cite{wilcox}
they can be determined as:
\begin{eqnarray}
    C_{2} & = & \frac{1}{2} [H_{I},H_{0}]
    \label{ZA}\\
    C_{3} & = & -\frac{1}{6} [H_{0},[H_{I},H_{0}]]-\frac{1}{3} [H_{I},[H_{I},H_{0}]]
    \label{Zasse1}
    \end{eqnarray}
with increasing complexity for higher $i$. The basic step in the
construction of approximants to the Eq.(\ref{Zasse}) is to find a
product of exponential operators which is correct up to a certain
power of $\beta$. The disentangled and undisentangled form of Eq.
(\ref{Zasse}) are expanded in terms of $\beta$ and operator
coefficients of equal power of $\beta$ are compared. The result is
that
\begin{equation}\label{nhao}
    e^{-\beta
    (H_{0}+H_{I})}=e^{-\beta {H_{0}}}e^{-\beta
    {H_{I}}}+O(\beta^2)\,,
\end{equation}
A more accurate approximation is obtained by introducing the
following symmetrized approximation of the Hermitian operator
$e^{-\beta H}$:
\begin{equation}\label{hao}
    e^{-\beta(H_{0}+H_{I})}=e^{-{\beta H_{0}}/{2}}e^{-{\beta H_{I}}}e^{-{\beta H_{0}}/{2}}+O(\beta^3)\,,
\end{equation}
Obviously, the partition function obtained with using (\ref{hao})
and (\ref{nhao}) is identical due to the cyclic permutation
property of the trace.
The error induced by the approximation
\begin{equation}\label{exprwa}
    {\mathrm{Tr}}\left\{e^{-\beta
    (H_{0}+H_{I})}\right\}={\mathrm{Tr}}\left\{e^{-\beta H_{0}}e^{-\beta H_{I}}\right\}
\end{equation}
can be estimated using the Hermitian approximation
\begin{eqnarray}\label{hao2}
    &e^{-\beta(H_{0}+H_{I})}\nonumber\\&=
    e^{-{\beta H_{0}}/{2}}e^{-{\beta H_{I}}/{2}}
    e^{-{\beta^{3} C_{3}}/{4}}e^{-{\beta H_{I}}/{2}}e^{-{\beta
    H_{0}}/{2}}+O(\beta^5)
\end{eqnarray}
and one obtains \cite{approx2}
\begin{eqnarray}
    &\left|{\mathrm{Tr}}\left\{e^{-\beta (H_{0}+H_{I})}-e^{-\beta H_{0}/2}e^{-\beta H_{I}}e^{-\beta
    H_{0}/2}\right\}\right|\nonumber\\&
    <\frac{\beta^3}{4}\left|{\mathrm{Tr}}\left\{C_3\right\}\right|
    \left|{\mathrm{Tr}}\left\{e^{-\beta H_{0}}e^{-\beta
    H_{I}}\right\}\right|
    \label{exprwaapp}
    \end{eqnarray}
\section{\label{modela}The Model}
We use this method to study the thermodynamic properties of the
Dicke model Hamiltonian ($\hbar=c=1$)
\begin{equation}
    H=\omega a^{\dagger}a + \sum_{i=1}^{N}\left[\frac{\epsilon}{2} \sigma_{i}^{z} +
    \frac{\lambda}{\sqrt{N}}(a^{\dagger}+a)(\sigma_{i}^{+}+\sigma_{i}^{-})\right]
    \label{hamrwa}
    \end{equation}
Here, $\omega$ is the frequency of a single mode of radiation,
$\epsilon$ is the energy difference between the two levels of N
identical spin-$\frac{1}{2}$ systems, $\sigma_{i}^{z}$,
$\sigma_{i}^{+}$ and $\sigma_{i}^{-}$ are respectively the $z$
component, the raising and the lowering operators of the Pauli
matrices used to describe the $i$th spin, $a$ and $a^{\dagger}$
are the annihilation and creation operators for photons. For N
two-level atoms the coupling constant is
\begin{equation}
    \lambda=\epsilon d \sqrt{\frac{2 \pi \rho} {\omega}}
    \label{parat}
    \end{equation}
where $d$ is the projection of the transition dipole moment on the
polarization vector of the field mode and $\rho$ the density of
the atoms.

We make the following separation:
\begin{eqnarray}\label{hammmm}
    H_{0}&=&\omega a^{\dagger}a,\\
    H_{I}&=&
    \frac{\epsilon}{2} S^{z}+\frac{\lambda}{\sqrt{N}}(a^{\dagger}+ a
    )(S^{+}+S^{-})
    \label{ham1}
    \end{eqnarray}
where
\begin{equation}\label{cao}
    S^{(z,\pm)}=\sum_{i=1}^{N}\sigma_{i}^{(z,\pm)}
\end{equation}
are the collective atomic operators. By applying the
harmonic-oscillator commutation relations of the field mode
operators $a^{\dagger}$, $a$:
\begin{equation}
    [a,a^{\dagger}]=1,\quad [a^{\dagger}a,a^{\dagger}]=a^{\dagger},\quad
    [a^{\dagger}a,a]=-a
    \label{comfield}
    \end{equation}
and that for the atomic operators:
\begin{equation}
    [S^{+},S^{-}]=S^{z},\quad
    [S^{z},S^{\pm}]=\pm 2 S^{\pm}
    \label{comatom}
    \end{equation}
it is easy to show that
\begin{eqnarray}
    C_{2}& =&\frac{\lambda\omega}{2\sqrt{N}}(a-a^{\dagger})(S^{-}+S^{+})
    \nonumber\\
    C_{3} & = &
    \frac{\lambda\omega}{6\sqrt{N}}\left[\omega(a+a^{\dagger})\left(S^{+}+S^{-}\right)+
    2\epsilon(a-a^{\dagger})\left(S^{-}-S^{+}\right)\right.\nonumber\\
    &+&\left.\frac{4\lambda}{\sqrt{N}}\left(S^{-}+S^{+}\right)^2\right]
    \label{comm1}
    \end{eqnarray}
In order to evaluate the quality of the approximation we calculate
the trace of $C_{3}$ that requires summation over both the atomic
and the field variables. A sum over atomic variables is
well-suited to calculating the trace, yielding
\begin{equation}\label{trac4}
    \sum_{S_1=\pm 1}\dots\sum_{S_N=\pm
    1}\langle S_1\dots S_N|C_3|S_1\dots
    S_N\rangle=\frac{4}{3}\lambda^2\omega
\end{equation}
The expression (\ref{exprwa}) is correct up to the third order in
$\beta$ (or rather for $\beta^3\lambda^2\omega<1$), i.e. is an
appropriate description for the high-temperature limit.

Although it is possible to derive higher-order approximations of
partition function in a systematic manner, the increasing
complexity of $C_i$ for higher $i$ now complicates the expressions
in such way that $Z(N,T)$ is very difficult to evaluate
analytically and requires numerical calculations \cite{approx3}.
It is important to note that also a good choice of decomposing the
Hamiltonian may affect the complexity of the calculation and the
error induced by the approximation (\ref{exprwa}). A different
separation was studied in Ref. \cite{liberti}, in the simplest
case of exact resonance between atom energy levels and frequency
of radiation ($\omega=\epsilon$). The approximations that one
obtains from this choice is derived splitting off from
(\ref{ham1}) the ${\epsilon} S^{z}/2$ term too and is, for this
reason, less accurate.
\section{\label{partition}Partition function: Atomic variables}
In the form of Eq. (\ref{exprwa}), the partition function may be
performed analytically. Writing out explicitly the trace over the
atomic variables, the partition function is given by
\begin{eqnarray}
     Z(N,T) &=& {\mathrm{Tr_{F}}}\left\{e^{-\beta\omega
     a^{\dagger}a}\sum_{S_1=\pm 1}\cdots\sum_{S_N=\pm 1}\right.\nonumber\\
    &\times&\left.\langle S_1,\dots,S_N|
    e^{-\beta\sum_{j=1}^N h_j}|S_1,\dots,S_N\rangle\right\}
    \label{Zeta}
\end{eqnarray}
where
\begin{equation}\label{defh}
    h_{j}=\frac{\epsilon}{2}\sigma_j^{z}+\frac{\lambda}{\sqrt{N}}
    (\sigma_j^{+}+\sigma_j^{-})(a^{\dagger}+ a)
\end{equation}
Noting that this operator has the property
\begin{equation}\label{comdefh}
    \left[h_{i},h_{j}\right]=0\,,\quad(i\neq
    j)
\end{equation}
from which it follows that
\begin{equation}\label{sumprod2}
    e^{-\beta\sum_{j=1}^N h_{j}}=\prod_{j=1}^N e^{-\beta
    h_j}
\end{equation}
we can reduce the partition function to the simpler form
\begin{eqnarray}
    Z(N,T) &=& {\mathrm{Tr_{F}}}\left\{e^{-\beta\omega a^{\dagger}a}\right.\nonumber\\
    &\times&\left.\left[\sum_{S=\pm 1}
    \langle S|e^{-\beta\left[\frac{\epsilon}{2}\sigma^{z}+\frac{\lambda}{\sqrt{N}}
    (a^{\dagger}+ a)\sigma^{x}\right]}|S\rangle\right]^N\right\}
    \label{Zetanew}
\end{eqnarray}
where $\sigma^x\equiv\sigma^++\sigma^-$.
Expanding the exponential operators in a power series, we obtain
\begin{eqnarray}\label{expexp}
    &&e^{-\beta\left[\frac{\epsilon}{2}\sigma^{z}+\frac{\lambda}{\sqrt{N}}
    (a^{\dagger}+
    a)\sigma^{x}\right]}=\sum_{k=0}^\infty\frac{\beta^{2k}}{(2k)!}
    \left[\frac{\epsilon^2}{4}+\frac{\lambda^2}{N}(a^{\dagger}+a)^2\right]^k
    \nonumber\\&&\times\left[I-\frac{\beta}{2k+1}\left(\frac{\epsilon}{2}\sigma^z+\frac{\lambda}{\sqrt{N}}
    (a^{\dagger}+a)\sigma^x\right)\right]
\end{eqnarray}
where we have used the following Pauli matrices properties
\begin{equation}\label{pauli}
    \sigma_z^2=\sigma_x^2=I\,,\quad\sigma^z\sigma^x+\sigma^x\sigma^z=0
\end{equation}
Therefore, the sum of Eq.(\ref{Zetanew}) is given by
\begin{eqnarray}\label{sumsvolta}
    &\sum_{S=\pm 1}\langle S|
    e^{-\beta\left[\frac{\epsilon}{2}\sigma^{z}+\frac{\lambda}{\sqrt{N}}
    (a^{\dagger}+ a)\sigma^{x}\right]}|S\rangle\nonumber\\
    &=2\sum_{k=0}^\infty
    \frac{\beta^{2k}}{(2k)!}\left[\frac{\epsilon^2}{4}+\frac{\lambda^2}{N}(a^{\dagger}+a)^2\right]^k\,.
\end{eqnarray}
\section{\label{field}Partition function: Field variables}
Using the Fock-state $|n \rangle$ for the photon field, the
partition function is given by
 \begin{eqnarray}
    &&Z(N,T) = 2^N\sum_{n=0}^\infty\exp{(-\beta\omega n)}\nonumber\\
    &\times&\langle n|\left\{\sum_{k=0}^\infty \frac{\beta^{2k}}{(2k)!}
    \left[\frac{\epsilon^2}{4}+\frac{\lambda^2}{N}(a^{\dagger}+a)^2\right]^k\right\}^N|n\rangle\,.
    \label{Zetaco}
\end{eqnarray}

Since the operator $(a^{\dagger}+a)$ commutes with itself, the
power series appearing into the above equation can be written
\begin{eqnarray}
    & &\left\{\sum_{k=0}^\infty
    \frac{\beta^{2k}}{(2k)!}\left[\frac{\epsilon^2}{4}+\frac{\lambda^2}{N}(a^{\dagger}+a)^2\right]^k\right\}^N
    =\left(\prod_{i=1}^N\sum_{k_i=0}^\infty\frac{\beta^{2k_i}}{(2k_i)!}\right)\nonumber\\
    &&\times
    \sum_{q=0}^K
    \frac{K!}{q!(K-q)!}\left(\frac{\epsilon}{2}\right)^{2(K-q)}\left(\frac{\lambda}{\sqrt{N}}\right)^{2q}
    (a^{\dagger}+a)^{2q}
    \label{Zetasumme}
\end{eqnarray}
where $K=k_1+\cdots+k_N$. The matrix elements of photon operators
are
\begin{eqnarray}\label{ma}
&&\langle
n|(a^{\dagger}+a)^{2q}|n\rangle=\left.\frac{d^{2q}}{d\eta^{2q}}\langle
n|e^{\eta(a^{\dagger}+a)}|n\rangle\right|_{\eta=0}\nonumber\\
&&=\frac{d^{2q}}{d\eta^{2q}}\left[e^{\frac{\eta^2}{2}}L_n(-\eta^2)\right]_{\eta=0}
\end{eqnarray}
where $L_n(x)$ is the $n$th Laguerre polynomial. At this point
 we are able to write down the partition function (\ref{Zetaco}) as
\begin{eqnarray}
    Z(N,T) &=&2^N\left(\prod_{i=1}^N\sum_{k_i=0}^\infty\frac{\beta^{2k_i}}{(2k_i)!}\right)\sum_{q=0}^K
    \frac{K!}{q!(K-q)!}\left(\frac{\epsilon}{2}\right)^{2(K-q)}
    \nonumber\\
    & \times&
    \left.\left(\frac{\lambda}{\sqrt{N}}\right)^{2q}\frac{d^{2q}}{d\eta^{2q}}e^{\frac{\eta^2}{2}}\sum_{n=0}^\infty e^{-\beta\omega
    n}L_n(-\eta^2)\right|_{\eta=0}
    \label{Zetacohancora}
\end{eqnarray}
The sum over $n$
is well known \cite{arfken} to be given by
\begin{equation}\label{lag2}
    \sum_{n=0}^\infty e^{-\beta\omega
    n}L_n(-\eta^2)=\frac{1}{1-e^{-\beta\omega}}
    \exp\left[\eta^2\frac{1}{e^{\beta\omega}-1}\right]
\end{equation}
Substituting this into  Eq. (\ref{Zetacohancora}) one obtains
\begin{eqnarray}
    Z(N,T) &=&\frac{2^N}{1-e^{-\beta\omega}}\left(\prod_{i=1}^N\sum_{k_i=0}^\infty\frac{\beta^{2k_i}}{(2k_i)!}\right)
     \sum_{q=0}^K
    \frac{K!}{q!(K-q)!}\nonumber\\&\times&
    \left(\frac{\epsilon}{2}\right)^{2(K-q)}\left(\frac{\lambda}{\sqrt{N}}\right)^{2q}
    \frac{d^{2q}}{d\eta^{2q}}\left.e^{\frac{\eta^2}{2}\coth{\frac{\beta\omega}{2}}}\right|_{\eta=0}
    \label{Zetacohancora2}
\end{eqnarray}
At this stage we use the following result
\begin{equation}\label{potexpip}
    \left.\frac{d^{2q}}{d\eta^{2q}}e^{\frac{\eta^2}{2}\coth{\frac{\beta\omega}{2}}}\right|_{\eta=0}
    =(2q-1)!!\coth^q{\frac{\beta\omega}{2}}\,,\quad q\geq0
\end{equation}
and the integral representation for the double factorial
\cite{arfken}
\begin{equation}\label{intrepr}
    (2q-1)!!=\frac{1}{2^{q+1}\sqrt{\pi}}\int_{-\infty}^\infty dz
    e^{-\frac{z^2}{4}}z^{2q}\,.
\end{equation}
So, one has
\begin{eqnarray}
     Z(N,T)
    &=&\frac{1}{1-e^{-\beta\omega}}\frac{2^N}{\sqrt{4\pi}}\int_{-\infty}^\infty dz
    e^{-\frac{z^2}{4}}\left(\prod_{i=1}^N\sum_{k_i=0}^\infty\frac{\beta^{2k_i}}{(2k_i)!}\right)\nonumber\\
    & \times&
    \sum_{q=0}^K
    \frac{K!}{q!(K-q)!}\left(\frac{\epsilon}{2}\right)^{2(K-q)}\left(\frac{\lambda z}{\sqrt{2N}}\right)^{2q}
   \coth^q{\frac{\beta\omega}{2}}\nonumber\\
    &
    =&\frac{1}{1-e^{-\beta\omega}}\frac{1}{\sqrt{4\pi}}\int_{-\infty}^\infty dz e^{-\frac{z^2}{4}}\nonumber\\
    &\times&
    \left\{2\sum_{k=0}^\infty \frac{\beta^{2k}}{(2k)!} \left[\frac{\epsilon^2}{4}+
    \frac{\lambda^2 z^2}{2N}\coth{\frac{\beta\omega}{2}}\right]^k\right\}^N
    \label{Zetacovolo}
\end{eqnarray}
which can be written in the final form
\begin{eqnarray}
    Z(N,T)&=&\frac{1}{1-e^{-\beta\omega}}\frac{1}{\sqrt{4\pi}}
    \int_{-\infty}^\infty dz
    e^{-\frac{z^2}{4}}\nonumber\\&\times&\left\{2\cosh{\left[\beta\sqrt{\frac{\epsilon^2}{4}+
    \frac{\lambda^2 z^2}{2N}\coth{\left(\frac{\beta\omega}{2}\right)}}\right]}\right\}^N
    \label{Zetalast}
\end{eqnarray}
where $z$ is an order parameter. This result is independent from
the choice of different states as a basis for expressing the state
of radiation field and the same result can be derived using the
coherent states $|\alpha\rangle$ for the photon field (see App.
\ref{A}).
\section{\label{phase}Phase transition: classical limit}
The integral (\ref{Zetalast}) may be evaluated in the limit
$N\rightarrow \infty$ by the steepest descent method. Writing the
integral in term of a new variable $x={z}/\sqrt{N}$ we search the
value $\tilde{x}$ for which
 \begin{equation}
      f(x)=- \frac{x^2}{4} +
      \ln\left\{2\cosh{\left[\beta\sqrt{\frac{\epsilon^2}{4}+\frac{\lambda^2 x^2}{2}
      \coth{\left(\frac{\beta\omega}{2}\right)}}\right]}\right\}
      \label{fx}
      \end{equation}
is minimized. The minimum condition implies
\begin{eqnarray}
    &&\beta\lambda^2\tanh{\left[\beta\sqrt{\frac{\epsilon^2}{4}+\frac{\lambda^2
    x^2}{2}\coth{\left(\frac{\beta\omega}{2}\right)}}\right]}\nonumber\\&&=
    \tanh{\left(\frac{\beta\omega}{2}\right)}\sqrt{\frac{\epsilon^2}{4}+\frac{\lambda^2
    x^2}{2}\coth{\left(\frac{\beta\omega}{2}\right)}}
    \label{f'x2}
    \end{eqnarray}
The existence of a nonzero solution of the above equation means a
phase transition. Through Eq. (\ref{f'x2}) we can compute a
critical temperature and we find
\begin{equation}\label{tempcrit}
    \beta_c=
    \frac{\epsilon}{2\lambda^2}\frac{\tanh{\left({\beta_c\omega}/{2}\right)}}
    {\tanh{\left({\beta_c\epsilon}/{2}\right)}}
\end{equation}
The quantity $\langle S_z\rangle$ is of physical interest. We
calculate it as follows:
\begin{equation}\label{am1}
    \langle
    S_z\rangle=-\frac{2}{\beta}\frac{\partial}{\partial\epsilon}\ln{Z(N,T)}
\end{equation}
From Eq. (\ref{f'x2}) we infer that
\begin{equation}\label{am2}
    \frac{\langle S_z\rangle}{N}=\left\{%
\begin{array}{ll}
    -\tanh{\left(\frac{\beta\epsilon}{2}\right)}, & \beta<\beta_c\,; \\
    -\frac{\epsilon}{2\beta\lambda^2}\tanh{\left(\frac{\beta\omega}{2}\right)}, & \beta>\beta_c\,. \\
\end{array}
\right.
\end{equation}
A special case of interest is the limit reached when
$\beta\omega\ll 1$. In this limit the partition function, which is
given by Eq. (\ref{Zetalast}), becomes
\begin{eqnarray}
    Z(N,T)&\simeq&\frac{1}{\sqrt{4\pi}\beta\omega}\int_{-\infty}^\infty dz
    e^{-\frac{z^2}{4}}\nonumber\\&\times&\left[2\cosh{\left(\beta\sqrt{\frac{\epsilon^2}{4}+\frac{\lambda^2
    z^2}{N\beta\omega}}\right)}\right]^N
    \label{Zetalastapp}
\end{eqnarray}
Let
\begin{equation}\label{wd}
    y^2=\frac{z^2}{4N\beta \omega}
\end{equation}
one has
\begin{eqnarray}
    Z(N,T)&\simeq&\sqrt{\frac{N}{\beta\omega{\pi}}}\int_{-\infty}^\infty
    dy
    e^{-{N\beta\omega y^2}}\nonumber\\&\times&\left[2\cosh{\left({\beta}\sqrt{\frac{\epsilon^2}{4}+{4\lambda^2
    y^2}}\right)}\right]^N
    \label{Zetalastapp2}
\end{eqnarray}
that correspond to the partition function obtained with the Wang
and Hioe computational method. In this limit, the square of the
order parameter $y$ represent the average number of photons. Just
as for the Wang and Hioe result \cite{Duncan}, the critical
temperature is obtained from the equation
\begin{equation}\label{tempcrit2}
    \tanh{\left(\frac{\beta_c\epsilon}{2}\right)}=\frac{\epsilon\,\omega}{4\lambda^2}
\end{equation}
and the model Hamiltonian (\ref{hamrwa}) undergoes a phase
transition at the critical value of the coupling constant
$\lambda_c={\sqrt{\epsilon\,\omega}}/{2}$. For a coupling
$\lambda>\lambda_c$, above the critical temperature, the system is
in the "normal phase", whereas for $\beta>\beta_c$, the equation
(\ref{f'x2}) has a solution $\tilde{x}\neq 0$ and the system is in
the so-called "super-radiant phase". In this region, the
expectation value of the collective angular momentum operator
(\ref{am2}) is
\begin{equation}\label{am3}
    \frac{\langle S_z\rangle}{N}=\left\{%
\begin{array}{ll}
    -\tanh{\left(\frac{\beta\epsilon}{2}\right)}, & \beta<\beta_c\,; \\
    -\frac{\epsilon\omega}{4\lambda^2}, & \beta>\beta_c\,. \\
\end{array}
\right.
\end{equation}
\section{\label{modelb}Phase transition: Ising limit}
The equivalence between the Dicke Model and the one-dimensional
Ising Model for a system of mutually interacting spins 1/2
embedded in a transverse magnetic field \cite{Parisi} has been
studied with different techniques \cite{lambert,reslen,liberti}.
This similarity emerges in our approximate scheme when
$\beta\epsilon\ll 1$, where Eq. (\ref{tempcrit}) becomes
\begin{equation}\label{tempcr}
    \tanh{\left(\frac{\beta_c\omega}{2}\right)}=\beta_c^2\lambda^2
\end{equation}
This result may be easily obtained observing that, in this limit,
our model is equivalent to a high-temperature expansion of the
temperature-dependent effective Hamiltonian given by
\begin{equation}\label{is1}
    H(\beta)=\omega\, a^\dagger
   a+\frac{\epsilon}{2}S_z-\frac{\beta\lambda^2}{2N}\coth{\left(\frac{\beta\omega}{2}\right)}S_x^2
\end{equation}
This effective Hamiltonian is the lowest-order effective
Hamiltonian that can be obtained using the method of Ref.
\cite{reslen} (see App. \ref{B}).
As an alternative to the expression of Eq. (\ref{Zetalast}), the
partition function can be written as
\begin{eqnarray}
    Z(N,T) &=&\frac{1}{1-e^{-\beta\omega}}\frac{1}{\sqrt{4\pi}}
    \int_{-\infty}^\infty dz
    e^{-\frac{z^2}{4}}\nonumber\\&\times&{\mathrm{Tr_{A}}}\left\{
    e^{-\beta\left[\frac{\epsilon}{2}S_z+\frac{\lambda z}{\sqrt{2N}}
    \sqrt{\coth{\left(\frac{\beta\omega}{2}\right)}}S_x\right]}\right\}
    \label{alt}
\end{eqnarray}
i.e., in the high temperature limit,
\begin{eqnarray}
    Z(N,T) &\simeq&
   \frac{1}{1-e^{-\beta\omega}}\frac{1}{\sqrt{4\pi}}\int_{-\infty}^\infty dz
    e^{-\frac{z^2}{4}}\nonumber\\&\times&{\mathrm{Tr_{A}}}\left\{e^{-\beta
    \frac{\epsilon}{2}S^{z}}
    e^{\frac{\beta\lambda z}{\sqrt{2N}}\sqrt{\coth{\left(\frac{\beta\omega}{2}\right)}}S_x}\right\}\nonumber\\
    &=& \frac{1}{1-e^{-\beta\omega}}{\mathrm{Tr_{A}}}\left\{e^{-\beta
    \frac{\epsilon}{2}S^{z}}
    e^{\frac{\beta^2\lambda^2}{2N}\coth{\left(\frac{\beta\omega}{2}\right)}S_x^2}\right\}
    \label{is3}
\end{eqnarray}
As in the most general case (\ref{Zetalast}), we can write the
integral in term of $x={z}/\sqrt{N}$ and utilize (in the limit
$N\rightarrow \infty$) the steepest descent method in order to
search the value $\tilde{x}$ for which
 \begin{eqnarray}
      f(x)&=&- \frac{x^2}{4} +
      \ln\left\{2\cosh{\left(\frac{\beta\epsilon}{2}\right)}\right.
      \nonumber\\ &\times&\left.
      \cosh{\left[\frac{\beta\lambda z}{\sqrt{2N}}
    \sqrt{\coth{\left(\frac{\beta\omega}{2}\right)}}\right]}\right\}
      \label{fxb}
      \end{eqnarray}
is minimized. The minimum condition implies
\begin{equation}
    \frac{\tilde{x}}{2}=\frac{\beta\lambda }{\sqrt{2}}
    \sqrt{\coth{\left(\frac{\beta\omega}{2}\right)}}\tanh{\left[\frac{\beta\lambda \tilde{x}}{\sqrt{2}}
    \sqrt{\coth{\left(\frac{\beta\omega}{2}\right)}}\right]}
    \label{f'x2b}
    \end{equation}
The existence of a nonzero solution of the above equation means a
phase transition. Through Eq. (\ref{f'x2b}) we can compute a
critical temperature that we find to be identical to the result
(\ref{tempcr}). The order parameter near the critical temperature
is
\begin{equation}\label{op}
    \tilde{x}\simeq
    2\sqrt{3}\;\sqrt{\frac{\beta-\beta_c}{\beta_c}}\,.
\end{equation}
 In the limit $\beta_{c}\omega\ll1$, we get
$\beta_{c}\simeq{\omega}/{2\lambda^2}$ and the interaction of
independent spins with a photon mode induce an effective spin-spin
interaction of long range nature that may be described by the
Hamiltonian
\begin{equation}\label{is4}
    H=\omega\, a^\dagger
    a-\frac{\lambda^2}{N\omega}S_x^2
\end{equation}
that is in agreement with the results of Ref.
\cite{Emary0,lambert}.

Before leaving this section, we note that, for the special case of
exact resonance $\omega=\epsilon$, Eq. (\ref{tempcrit}) reduces to
\begin{equation}\label{tcr}
    \beta_c=\frac{\epsilon}{2\lambda^2}
\end{equation}
and this lead to the absence of a critical value of the coupling
constant, i.e. to the result that the phase transition could occur
even for $\lambda<\epsilon/2$. However, the approximation that we
have proposed is valid for $\beta_c^3\lambda^2\omega<1$ and this
condition permit us to derive accurate results only for
$\lambda>\epsilon/2$, i.e. for coupling sufficiently higher that
$\lambda_c$.
\section{\label{conclu}Conclusion}
The thermodynamic properties of a system governed by the Dicke
Hamiltonian have been treated in the framework of an approximate
model, through a perturbative expansion of partition function
obtained by decomposing the Hamiltonian into two non-commuting
hermitian operators. This technique represent a practical and
convenient method to determinate the behavior of the Dicke model
in the strong-coupling regime. We have obtained a simple analytic
expression for the partition function and the critical temperature
is easily determined. This technique was then used to explore some
limiting cases and old and recent results are derived in an
elementary and unified way. Our result extend those of Ref.
\cite{liberti} beyond the resonant interaction condition and
support the recent arguments on the subject \cite{lambert,reslen}
concerning the similarity between the Dicke and collective
one-dimensional Ising model.
\appendix
\section{\label{A}Coherent states}
In terms of coherent states $|\alpha\rangle$ \cite{glauber}, the
partition function is given by
\begin{eqnarray}
    Z(N,T) &=& 2^N\int\frac{d^2 \alpha}{\pi}
    \langle \alpha|e^{-\beta\omega a^{\dagger}a}\nonumber\\
    &\times&\left\{\sum_{k=0}^\infty \frac{\beta^{2k}}{(2k)!}
    \left[\frac{\epsilon^2}{4}+\frac{\lambda^2}{N}(a^{\dagger}+a)^2\right]^k\right\}^N|\alpha\rangle
    \label{a1}
\end{eqnarray}
where $a|\alpha\rangle=\alpha|\alpha\rangle$ and
\begin{equation}\label{a2}
    \int\frac{d^2 \alpha}{\pi}|\alpha\rangle\langle\alpha|=1.
\end{equation}
Using the over-completeness of the field coherent states
(\ref{a2}) and the result
\begin{equation}
    \langle \alpha|e^{-\beta\omega a^{\dagger}a}|\gamma\rangle =
    \langle \alpha|\gamma\rangle e^{-\alpha^{\ast}\gamma
    [1-\exp(-\beta\omega)]}
    \label{a3}
    \end{equation}
Eq. (\ref{a1}) takes the form
\begin{eqnarray}
    Z(N,T) &=& 2^N\int\frac{d^2 \alpha}{\pi}\int\frac{d^2 \gamma}{\pi}\langle \alpha|\gamma\rangle e^{-\alpha^{\ast}\gamma
    [1-\exp(-\beta\omega)]}\nonumber\\
    &\times&\langle\gamma|\left\{\sum_{k=0}^\infty \frac{\beta^{2k}}{(2k)!}
    \left[\frac{\epsilon^2}{4}+\frac{\lambda^2}{N}(a^{\dagger}+a)^2\right]^k\right\}^N|\alpha\rangle\,.
    \label{a4}
\end{eqnarray}
The partition function may be derived from Eq. (\ref{Zetasumme})
by computing the matrix element
\begin{eqnarray}\label{a5}
    \langle \gamma|(a^{\dagger}+a)^{2q}|\alpha\rangle&=&\left.\frac{d^{2q}}{d\eta^{2q}}\langle
    \gamma|e^{\eta(a^{\dagger}+a)}|\alpha\rangle\right|_{\eta=0}\nonumber\\&=&
    \langle \gamma|\alpha\rangle\left.\frac{d^{2q}}{d\eta^{2q}}e^{\frac{\eta^2}{2}}e^{\eta(\gamma^{\ast}+\alpha)}\right|_{\eta=0}
\end{eqnarray}
and by using
\begin{equation}\label{a7}
    |\langle\alpha|\gamma\rangle|^2=e^{-|\alpha-\gamma|^2}\,.
\end{equation}
One obtains
\begin{eqnarray}
    &&Z(N,T) =2^N\left(\prod_{i=1}^N\sum_{k_i=0}^\infty\frac{\beta^{2k_i}}{(2k_i)!}\right)\nonumber\\
    & \times&\sum_{q=0}^K
    \frac{K!}{q!(K-q)!}\left(\frac{\epsilon}{2}\right)^{2(K-q)}\left(\frac{\lambda}{\sqrt{N}}\right)^{2q}\frac{d^{2q}}{d\eta^{2q}}e^{\frac{\eta^2}{2}}
    \nonumber\\
    & \times&
    \left.\int\frac{d^2 \alpha}{\pi} \int\frac{d^2 \gamma}{\pi}
    e^{-|\alpha|^2-|\gamma|^2+\gamma^\ast\alpha+\alpha^{\ast}\gamma
    e^{-\beta\omega}+\eta(\gamma^{\ast}+\alpha)}\right|_{\eta=0}
    \label{a6}
\end{eqnarray}
Recalling that the integration measure is defined to be given by
\begin{equation}\label{a8}
    \frac{d^2
    \alpha}{\pi}=\frac{{d\alpha}{d\alpha^\ast}}{2i}=\frac{{d({\mathrm{Re}}\alpha)}\,{d(\mathrm{Im}}\alpha)}{\pi}
\end{equation}
one finds the partition function expression of
Eq.(\ref{Zetacohancora2}), i.e. the same result obtained with the
use of Fock states as a basis for the photon field. Finally, we
want to underline that our results are derived without the
$c$-number substitution for the field variables
$a\rightarrow\alpha$, the Wang and Hioe computational method, that
permits to deal with non-interacting atoms subjected to an
external magnetic field described by the amplitude $\alpha$.
\section{\label{B}Effective Hamiltonian}
In this Appendix we will discuss the approach to the problem of
constructing effective atomic Hamiltonian for the Dicke Model. The
Zassenhaus formula (\ref{Zasse}) can be used to obtain the
following result:
\begin{equation}
    e^{-\beta
    H_{I}}\prod_{i=2}^\infty e^{(-\beta)^{i}
    C_{i}}=\sum_{i=0}^\infty \frac{(-\beta)^{i}}{i!} P_{i}
    \label{b1}
    \end{equation}
where
\begin{equation}\label{b2}
    P_{i}={H_I^i}+\sum_{k=2}^i
    \frac{i!}{(i-k)!}H_I^{i-k}C_k\,.
\end{equation}
The partition function of the whole system can be written as
\cite{approx3}:
\begin{eqnarray}\label{b3}
    Z(N,T)&=&{\mathrm{Tr_{A}}}\left[{\mathrm{Tr_{F}}}\left(e^{-\beta
    H_{0}}\sum_{i=0}^\infty \frac{(-\beta)^{i}}{i!}
    P_{i}\right)\right]\nonumber\\&=&{\mathrm{Tr_{F}}}\left(e^{-\beta H_0}\right){\mathrm{Tr_{A}}}\left(e^{-\beta H_A^{eff}}\right)
\end{eqnarray}
where
\begin{eqnarray}\label{b4}
    H_A^{eff}&=&-\frac{1}{\beta}\ln \left\langle\sum_{i=0}^\infty \frac{(-\beta)^{i}}{i!}
    P_{i}\right\rangle_F\nonumber\\&=&\frac{1}{\beta}\sum_{q=1}^\infty \frac{(-1)^q}{q}\left(\sum_{i=1}^\infty \frac{(-\beta)^{i}}{i!}
    \langle P_{i}\rangle_F\right)^q
\end{eqnarray}
where the thermal averaging is carried out with respect to the
field variables, i.e. $\langle
O\rangle_F={\mathrm{Tr_{F}}}\left(e^{-\beta
H_{0}}O\right)/{\mathrm{Tr_{F}}}\left(e^{-\beta H_{0}}\right)$.
Eq. (\ref{b4}) can be rewritten as
\begin{equation}\label{b5}
    H_A^{eff}=\sum_{q=1}^\infty \frac{\beta^{q-1}}{q!} Q_q
\end{equation}
where the lowest-order terms are
\begin{eqnarray}\label{b6}
    Q_1&=&\langle P_1\rangle_F\equiv\langle
    H_I\rangle_F=\frac{\epsilon}{2}S_z\,,\nonumber\\
    Q_2&=&\langle P_1\rangle_F^2-\langle P_2\rangle_F\equiv\langle H_I\rangle_F^2-\langle
    H_I^2\rangle_F\nonumber\\&=&
    -\frac{\lambda^2}{N}\coth{\left(\frac{\beta\omega}{2}\right)}S_x^2\,.
\end{eqnarray}
Therefore the lowest-order effective Hamiltonian is
\begin{equation}\label{b8}
H_A^{eff}=\frac{\epsilon}{2}S_z-\frac{\beta\lambda^2}{2N}\coth{\left(\frac{\beta\omega}{2}\right)}S_x^2
\end{equation}
in agreement with the result of Ref. \cite{reslen}.
\section{\label{C}Inclusion of diamagnetic term}
In this Appendix we briefly discuss the inclusion of diamagnetic
effects in our model. Following the discussion of Ref.
\cite{crisp} and \cite{liberti}, the diamagnetic term may be
incorporated in the Dicke Hamiltonian of Eq. (\ref{hamrwa}),
without adding significant complication of the problem, by using a
new field-mode of frequency $\Omega$ instead of frequency
$\omega$, which is given by
\begin{equation}
    \Omega=\sqrt{\omega (\omega + 4 k)}
    \label{nu}
    \end{equation}
where $\omega k=e^2\pi\rho/m$. In terms of this new frequency, the
model Hamiltonian becomes
\begin{equation}
    H=\Omega a^{\dagger}a + \sum_{i=1}^{N}\left[\frac{\epsilon}{2} \sigma_{i}^{z} +
    \frac{\Lambda}{\sqrt{N}}(a^{\dagger}+a)(\sigma_{i}^{+}+\sigma_{i}^{-})\right]
    \label{hamcrisp}
    \end{equation}
where
\begin{equation}
    \Lambda=\epsilon d \sqrt{\frac{2 \pi \rho} {\Omega}}
    \label{paratLa}
    \end{equation}
The Hamiltonian (\ref{hamcrisp}) is formally identical to the
Hamiltonian (\ref{hamrwa}) and can be used to derive thermodynamic
results. The phase transition that is obtained in the limit of
high coupling constant
($\Lambda\gg\Lambda_c=\sqrt{\epsilon\Omega}/2$) cannot occur due
to sum-rule arguments \cite{birula} that requires
$\Lambda<\sqrt{\epsilon\Omega}/2$.

\end{document}